\documentclass[final,5p,twocolumn,times,11pt]{elsarticle}
\usepackage{graphicx}  
\usepackage{color}     
\usepackage{epsfig}    
\usepackage{bm}        
\usepackage{amssymb}   

\advance\voffset -0.2in
\biboptions{sort&compress}
\newcommand{\beq}{\begin{equation}}
\newcommand{\eeq}{\end{equation}}
\newcommand{\beqar}{\begin{eqnarray}}
\newcommand{\eeqar}{\end{eqnarray}}
\newcommand{\ds}{\displaystyle}
\begin{document}

\begin{frontmatter}

\title{On the origin of forward-backward multiplicity correlations
       in $pp$ collisions at ultrarelativistic energies}

\author{L.V.~Bravina\corref{cor1}$^1$}
\ead{larissa.bravina@fys.uio.no}
\cortext[cor1]{Corresponding author}

\author{J.~Bleibel$^{2,3}$, E.E.~Zabrodin$^{1,4,5}$}

\address{$^1$ Department of Physics, University of Oslo, PB 1048 
              Blindern, N-0316 Oslo, Norway \\
$^2$ Institut f\"ur Angewandte Physik, Universit\"at T\"ubingen,
     Auf der Morgenstelle 10, D-72076 T\"ubingen, Germany \\
$^3$ Max-Planck-Institut f\"ur Intelligente Systeme,
     Heisenbergstr. 3, D-70569 Stuttgart, Germany \\
$^4$ Skobeltzyn Institute for Nuclear Physics,
     Moscow State University, RU-119899 Moscow, Russia \\
$^5$ National Research Nuclear University "MEPhI" (Moscow Engineering 
     Physics Institute), \\
     Kashirskoe highway 31, RU-115409 Moscow, Russia }

\date{\today}

\begin{abstract}
We study multiplicity correlations of hadrons in forward and backward 
hemispheres in $pp$ inelastic interactions at energies 200~GeV $\leq
\sqrt{s} \leq$ 13~TeV within the microscopic quark-gluon string model.
The model correctly describes (i) the almost linear dependence of 
average multiplicity in one hemisphere on the particle multiplicity in
other hemisphere in the center-of-mass frame; (ii) the increase of the 
slope parameter $b_{corr}$ 
with rising collision energy; (iii) the quick falloff of the correlation
strength with increase of the midrapidity gap; (iv) saturation of the
forward-backward correlations at very high multiplicities. Investigation
of the sub-processes on partonic level reveals that these features can be
attributed to short-range partonic correlations within a single string
and superposition of several sub-processes containing different numbers 
of soft and hard Pomerons with different mean multiplicities. If the 
number of Pomerons in the event is fixed, no forward-backward 
correlations are observed. Predictions are made for the top LHC energy 
$\sqrt{s} = 13$~TeV.
\end{abstract}



\end{frontmatter}

\section{Introduction}
\label{sec1}

The study of correlations between particles, produced in hadronic or 
nuclear collisions at high energies, can help us to reveal dynamical
features of multiparticle production \cite{WDK_96,Koch_10}. Among the 
first correlations measured in hadronic interactions, primarily 
(anti)proton$-$proton ones, were the multiplicity correlations between 
hadrons emitted in forward and in backward hemispheres, respectively, in 
the center-of-mass frame of the reaction. These corrections were 
extensively studied, e.g., by the UA5 Collaboration in $\bar{p}p$ 
collisions at ISR energies from $\sqrt{s} = 200$~GeV to 900~GeV 
\cite{ua5_1,ua5_2,ua5_3}. The linear dependence of the average 
multiplicity of particles emitted in forward hemisphere $\langle n_F 
\rangle$ on the particle multiplicity in backward hemisphere $n_B$, and 
vice versa, has been reported in \cite{uhlig_78}
\beq \ds
\langle n_F (n_B) \rangle = a + b_{corr} n_B\ ,
\label{eq1}
\eeq
where
\beq \ds
b_{corr} = \frac{\langle n_F n_B \rangle - \langle n_F \rangle^2 }
    {\langle n_F^2 \rangle - \langle n_F \rangle^2}
\label{eq2}
\eeq
Later on the linear dependence of the forward-backward (FB) correlations 
was also observed in $pp$ and $\bar{p}p$ collisions at lower and higher
energies, varying from $p_{lab} = 32$~GeV/$c$ ($\sqrt{s} = 7.86$~GeV)
\cite{fb_qgsm_89} and $p_{lab} = 250$~GeV/$c$ ($\sqrt{s} = 22$~GeV)
\cite{na22_89} to c.m. energies $0.3 \leq \sqrt{s} \leq 1.8$~TeV
\cite{e735}. At Large Hadron Collider (LHC), ALICE Collaboration has
confirmed the linear rise of $\langle n_B \rangle$ with increasing $n_F$ 
in $pp$ collisions at $\sqrt{s} = 0.9$, 2.76 and 7~TeV 
\cite{fb_alice_15}. Note that the distribution $\langle n_F (n_B) 
\rangle$ deviates from linear behaviour in the range of high particle 
multiplicities. In $e^+e^-$ annihilation, in contrast, only very weak FB 
correlations were found \cite{tasso_89,opal_94}. These peculiarities 
attract significant attention of theorists; see, e.g., 
\cite{CK_78,CV_82,DdD_81,CY_84,Car_88,Bar_87,CKC_97,Am_94,P_qm99,BPV_00,Vec_15}
and references therein.

Nowadays, the interest to the correlation phenomena in hadronic
interactions is raised because of the search for collective phenomena,
such as anisotropic flow, in these collisions. In the case of heavy ion
collisions, superposition of the first flow harmonics can explain the
characteristic long range correlations in rapidity space colloquially
known as {\it ridge} (see \cite{ALICE_ridge,ATLAS_ridge} and references 
therein).
Similar structure was found in high multiplicity $pp$ collisions as well
\cite{proton_ridge}. It remains an open question still whether or not
the ridge in $pp$ interactions is an initial-state or rather final-state
(i.e., collective) effect. The study of the FB correlations can shed
light on the origin of long range correlations emerging in hadronic
collisions at ultrarelativistic energies.

Application of quark-gluon string model (QGSM) 
\cite{qgsm_1,qgsm_2,qgsm_3,qgsm_mc1,qgsm_mc2} for 
the description of FB correlations in $pp$ and $\bar{p}p$ collisions at
$p_{lab} = 32$~GeV/$c$ \cite{fb_qgsm_89} was probably the first attempt
to uncover the role of different sub-processes of particle production
in the formation of FB correlations within the Monte Carlo (MC) 
microscopic model. Subsequently, these correlations were studied in $pp$
and $\bar{p}p$ collisions at higher energies within several microscopic 
models \cite{Am_94,dpm,paciae_10,WS_11,KV_14}. What about the 
quark-gluon string model? Obviously, the set of diagrams describing the 
variety of partonic sub-processes at hundreds and thousands GeV differs
from that corresponding to ten GeV. The relative contribution of each 
diagram also depends on the collision energy. Therefore, our present
study is focused on $pp$ interactions in the energy range $0.9 \leq
\sqrt{s} \leq 13$~TeV. A brief description of the basic principles of 
QGSM is given in Sec.~\ref{sec2}. The forward-backward multiplicity
correlations are studied in Sec.~\ref{sec3}. In particular, we 
demonstrate absence of the FB correlations in each sub-class of events
and appearance of strong positive correlations within the whole event
sample. Obtained results are also compared with the available 
experimental data. Conclusions are drawn in Sec.~\ref{sec4}.

\section{Quark-gluon string model}
\label{sec2}
    
Both the quark-gluon string model \cite{qgsm_1,qgsm_2,qgsm_3} and the 
dual parton model \cite{dpm} are based on the $1/N$ series expansion 
\cite{tH_74,Ven_74} of the amplitude for a QCD process. This approach is 
also known as {\it topological expansion\/}. For high energy processes 
with small momentum transfer there is a one-to-one correspondence 
between the diagrams arising in the topological expansion and the graphs 
corresponding to the exchange of Regge singularities, Reggeons and
Pomerons, in the Reggeon field theory (RFT) \cite{RFT_1,RFT_2}. Thus, 
the QGSM obeys the requirements of unitarity and analyticity. The 
amplitudes of multiparticle processes are found by cutting the diagrams 
in the $s$-channel. This procedure leads to formation of new objects, 
quark-gluon strings, which fragment into new hadrons and resonances
during the breakup stage.

\begin{figure}[htb]
 \resizebox{\linewidth}{!}{
\includegraphics{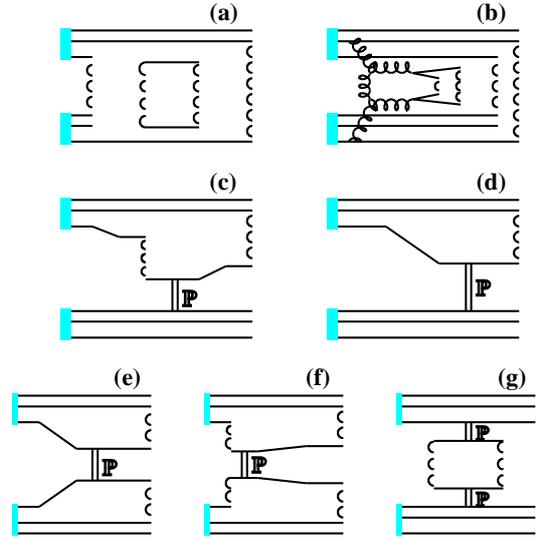}
}
\caption{
Diagrams taken into account in QGSM in the modeling of $pp$ interactions 
at ultrarelativistic energies: (a) multi-Pomeron exchange, (b) (semi)hard
gluon-gluon interaction and soft Pomeron exchange, (c)-(d) single 
diffraction with high-mass and low-mass excitation, (e)-(f) double
diffraction with low-mass and high-mass excitation, (g) central 
diffraction. }
\label{fig1}
\end{figure}
The Monte Carlo version of QGSM used for present calculations has been
employed for the description of hadronic, nuclear and hadron-nucleus
interactions, e.g., in 
\cite{fb_qgsm_89,qgsm_mc1,qgsm_mc2,ASC_92,FO_sps,v2_plb_01,prd_16}. 
Although the model contains a rich set of pre-asymptotic diagrams for 
hadron-hadron collisions at intermediate energies, at ultrarelativistic 
energies its number significantly shrinks. Here the total cross section 
of $pp$ interaction consists of just few terms, namely 
\beq \ds
\sigma_{tot}^{pp}(s) = \sigma_{el} + 
                    \sigma_P(s) + \sigma_{SD}(s) + \sigma_{DD}(s)\ ,
\label{eq3}
\eeq
where the terms in the rhs represent the cross sections of elastic, 
(multi)Pomeron, single-diffractive and double-diffractive processes,
respectively. Diagrams corresponding to last three processes are shown 
in Fig.~\ref{fig1}. To find the cross sections of single-diffractive and
double-diffractive processes displayed in Fig.~\ref{fig1}(c)-(g) we use
the parameterization \cite{prd_16} of the QGSM calculations made in
\cite{KP_11} 
\beqar \ds
\label{eq4}
\sigma_{SD}(s) &=& 0.68\, \left( 1 + 36\,s^{-1} \right)\,
\ln{ (0.6 + 0.2\,s) }\ , \\
\label{eq5}
\sigma_{DD}(s) &=& 1.65 + 0.27\,\ln{s}
\eeqar
It follows from the Froissart bound $\sigma_{tot} \propto (\ln{s})^2$
that the cross section of the diffractive processes should rise as
$\sigma_D \propto \ln{s}$.  Indeed, both $\sigma_{SD}(s)$ and
$\sigma_{DD}(s)$ in Eqs.~(\ref{eq4}),(\ref{eq5}) obey this asymptotic 
relation.

At ultrarelativistic energies, the processes going via exchange of soft 
Pomerons, shown in Fig.~\ref{fig1}(a), should be completed by the hard
Pomeron exchanges which lead to formation of hadronic jets, see 
Fig.~\ref{fig1}(b). This feature is attributed to all RFT-based MC 
models designed for the description of ultrarelativistic hadron-hadron 
and heavy-ion collisions \cite{dpm,ASC_92,epos,phojet,qgsjet2}. By
means of the Abramovskii-Gribov-Kancheli (AGK) cutting rules \cite{AGK}
the inelastic nondiffractive cross section is expressed as
\beq \ds
\nonumber
\sigma_{ND} (s) = \sum \limits_{i,j = 0; i+j \geq 1}^{ }
\sigma_{ij}(s)\ ,
\label{eq6}
\eeq
where
\beqar
\ds
\sigma_{ij}(s) &=& 2 \pi \int \limits_{0}^{\infty} b db\,
\exp{\left[ -2 u^R(s,b) \right]}\\
\nonumber
 &\times & \frac{\left[ 2u^R_{soft}(s,b)  \right] ^i}{i !}
\frac{\left[ 2u^R_{hard}(s,b)  \right] ^j}{j !} \ .
\label{eq7}
\eeqar
Here $b$ is the impact parameter, and superscript $R$ indicates real 
part of the eikonal $u(s,b) = u_{soft}(s,b) + u_{hard}(s,b)$. According
to \cite{CTVK_87}, both soft and hard eikonals can be cast in a form
\beqar \ds
\label{eq8}
u_{s/h}^R(s,b) &=& z_{s/h}(s)\, \exp{\left[- \frac{\beta^2}
{4\, \lambda_{s/h}(s)} \right]} \ , \\ 
\label{eq9}
z_{s/h}(s) &=& \frac{\gamma_P}{\lambda_{s/h}(s)}\,
\left( \frac{s}{s_0} \right)^\Delta \\
\label{eq10}
\lambda_{s/h}(s) &=& R_P^2 + \alpha_P^\prime \ln{\left(
\frac{s}{s_0} \right)} \ .
\eeqar
The quantities $\gamma_P$ and $R_P$ are Pomeron-nucleon vertex 
parameters, $\alpha_P (0)$ and $\alpha_P^\prime$ are the intercept and 
the slope of the Pomeron trajectory, respectively, $\Delta \equiv 
\alpha_P - 1$, and $s_0$ is a scale parameter. Numerical values of the 
principal parameters are listed in Table~\ref{tab1}.
The relative contributions of soft and hard Pomerons to particle spectra 
are energy dependent. At LHC energies the number of hard Pomerons per a
single $pp$ collision significantly increases, thus making the 
$p_T$-spectrum of secondaries harder, in full accord with the 
experimental data, see \cite{prd_16}.
\begin{table}
\caption{
\label{tab1}
Parameters of the soft and hard Pomerons used in the current version
of the QGSM.}
\vspace*{1ex}
\begin{tabular}{ccc}
\hline \hline
Parameter         & Soft Pomeron & Hard Pomeron \\
\hline \hline
$\gamma_P$        &    1.27475    &  0.021       \\
$R_P$             &    2.0        &  2.4         \\
$\alpha_P (0)$    &    1.15615    &  1.3217      \\
$\alpha_P^\prime$ &    0.25       &  0           \\
\hline
\end{tabular}
\end{table}
At very high energies one has to take into consideration Pomeron-Pomeron 
interactions described by the enhanced diagrams 
\cite{enh_diag_1,enh_diag_2}, which may cause the violation of the AGK 
cutting rules \cite{AGK-viol1,AGK-viol2}. However, these diagrams are 
not implemented in the current version of MC QGSM.

Finally, the Field-Feynman algorithm \cite{FF_frag} is employed to 
describe the fragmentation of strings with given quark content, mass and 
momentum into hadrons. The breakup of a string proceeds from the both 
edges with equal probabilities under the requirement of energy-momentum 
conservation and preservation of the quark content. Since no 
string-string interaction is taken into account, QGSM reveals only 
short-range correlations of hadrons in the rapidity space. 
Further details of the model can be found elsewhere 
\cite{qgsm_1,qgsm_2,qgsm_3,qgsm_mc1,qgsm_mc2,prd_16}.

\section{Results and discussion}
\label{sec3}
    
\begin{figure}[htb]
 \resizebox{\linewidth}{!}{
\includegraphics{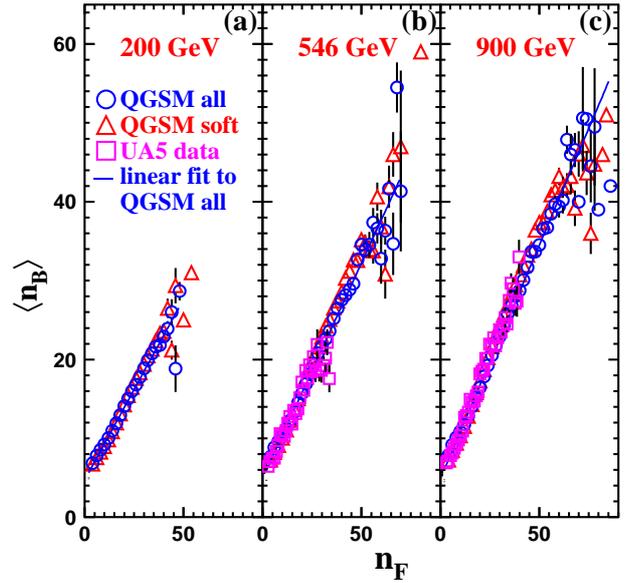}
}
\caption{
Distributions $\langle n_B \rangle (n_F)$ for
rapidity interval $0 \leq \eta \leq 4$ in NSD $pp$ collisions at (a) 
$\sqrt{s} = 200$~GeV, (b) 546~GeV and (c) 900~GeV. Open triangles
and circles indicate contributions of soft processes and all
processes, respectively. Data from \cite{ua5_2} are shown by open 
squares.
}
\label{fig2}
\end{figure}

QGSM successfully describes FB multiplicity correlations in hadronic 
interactions at energies below $\sqrt{s} = 100$~GeV 
\cite{fb_qgsm_89,na22_89}.
To examine the energy dependence of forward-backward correlations in 
$pp$ collisions at higher energies, the interactions were generated at 
$\sqrt{s} =  200$, 546 and 900~GeV. For each energy ca. one million 
non-single diffractive events were selected. Average multiplicities of 
charged hadrons emitted in backward hemisphere as function of charged 
particle yield in forward hemisphere in non-single diffractive (NSD) 
$pp$ collisions are displayed in Fig.~\ref{fig2} 
together with the available experimental results of UA5 \cite{ua5_2} 
collaboration. Recall, that the cross section of $\bar{p}p$ annihilation 
process drops quickly with rising collision energy, thus making possible 
comparison of $pp$ calculations with the $\bar{p}p$ data. To clarify the
role of hard processes, Fig.~\ref{fig2} shows separately the FB
correlations extracted from the model calculations with and without the 
hard Pomeron exchanges. The full distributions are fitted also to linear 
dependence given by Eq.~(\ref{eq1}). 

It is worth mentioning several features. (i) The shapes of the 
$\langle n_B \rangle (n_F)$ distributions are almost linear, except for 
very high $n_F$ multiplicities. (ii) The slope parameter $b_{corr}$ 
increases with rising $\sqrt{s}$ from $0.46 \pm 0.01$ at $\sqrt{s} = 
200$~GeV to $0.59 \pm 0.02$ at $\sqrt{s} = 900$~GeV. (iii) QGSM 
calculations agree well with the UA5 data at 546~GeV and 900~GeV. (iv)
There is no significant difference in FB multiplicity correlations 
between the model calculations with and without the hard processes, i.e. 
the forward-backward correlations arise mainly due to the soft 
processes. But, as was mentioned in Sec.~\ref{sec2}, the model possesses 
only the short-range correlations due to the dynamics of string 
break-up. We have to scrutinize, therefore, the FB correlations arising 
in the sub-processes of multiparticle production in QGSM. 

\begin{figure}[htb]
\vspace*{-1.2cm}
 \resizebox{\linewidth}{!}{
\includegraphics[scale=0.54]{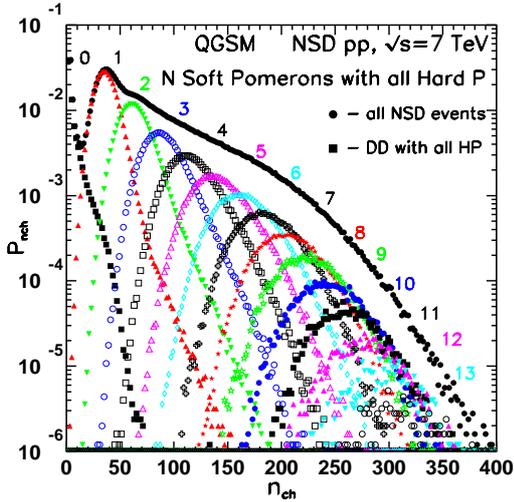}
}
\vspace*{-1.0cm}
\caption{
Multiplicity distributions of charged hadrons in NSD $pp$ collisions at 
$\sqrt{s} = 7$~TeV. Full circles and full squares denote all events and 
double-diffractive events together with all hard Pomerons, respectively. 
Other distributions are for the processes with 1,2, etc. soft Pomerons 
and all hard Pomerons.}
\label{fig3}
\end{figure}

\begin{figure}[htb]
\vspace*{-1.5cm}
 \resizebox{\linewidth}{!}{
\includegraphics{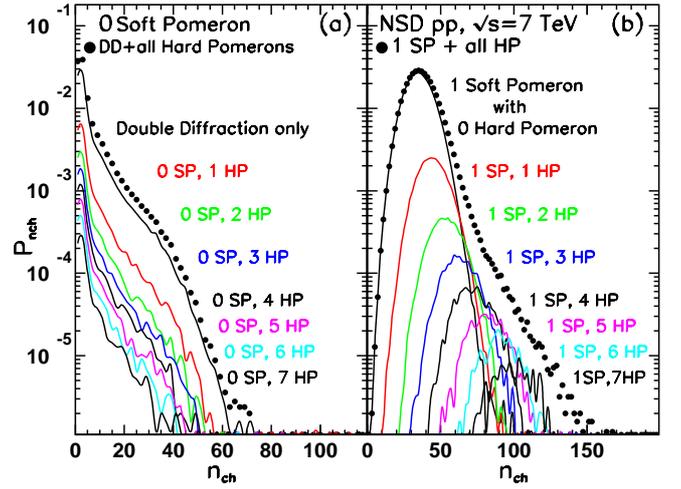}
}
\vspace*{-0.5cm}
\caption{
Multiplicity distributions of charged hadrons in NSD $pp$ collisions at 
$\sqrt{s} = 7$~TeV for processes with (a) zero and (b) one soft Pomerons. 
Solid circles show the total distributions, whereas lines indicate 
partial contributions of sub-events with 0, 1, 2, etc. hard Pomerons. }
\label{fig4}
\end{figure}

\subsection{Forward-backward multiplicity correlations in different 
sub-processes}
\label{subsec3.1}
    
Here $4 \cdot 10^6\ pp$ interactions at $\sqrt{s} = 7$~TeV are 
considered in order to
compare then the model calculations with the extensive data obtained at
this energy by ALICE collaboration \cite{fb_alice_15}. The multiplicity 
distributions of charged particles $P_{n_{ch}} (n_{ch})$ for NSD events 
with $i = 1,2, \ldots, 11$ soft Pomerons are shown in Fig.~\ref{fig3}. 
These distributions have a Gaussian-like structure and become broader 
with the increase of the number of soft Pomerons $N_{s\_Pom}$ in the 
event. The average multiplicity of charged hadrons also increases with 
rising $N_{s\_Pom}$. Note, that sub-processes with hard Pomerons are 
included in the analysis.  

To examine the role of hard processes, Figs.~\ref{fig4}(a) and 
\ref{fig4}(b) depict $P_{n_{ch}} (n_{ch})$ distributions for the events 
with zero and with one soft Pomeron, respectively, and up to seven hard 
Pomerons. In events without the soft Pomerons, the hard Pomerons enhance 
the multiplicity distribution of double-diffractive process by ca. 50\%, 
see Fig.~\ref{fig4}(a). The width of the final distribution, however, is 
almost insensitive to the number of hard Pomerons, $N_{h\_Pom}$. In 
contrast, events with one soft and several hard Pomerons, shown in 
Fig.~\ref{fig4}(b), demonstrate a shift in the positions of maxima of 
multiplicity distributions from 40 for 1SP+1HP events to 100 for 1SP+7HP 
ones. Similar effect takes place in events with two and more soft 
Pomerons. It causes the rise of high-multiplicity tails of the 
distributions shown in Fig.~\ref{fig3}.   

\begin{figure}[htb]
\vspace*{-1.0cm}
 \resizebox{\linewidth}{!}{
\includegraphics{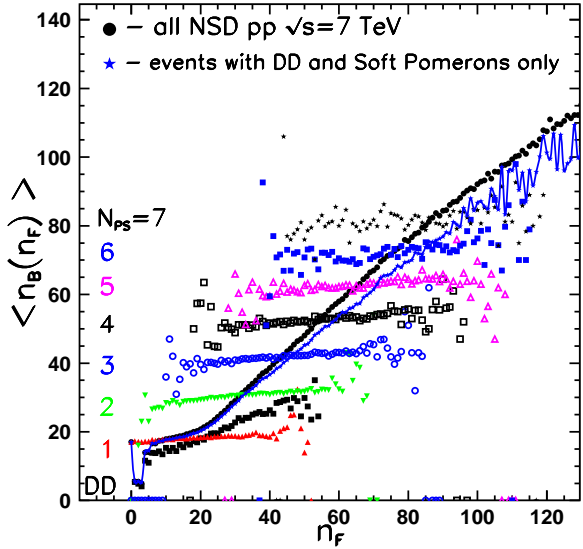}
}
\caption{
Dependence of average multiplicity $\langle n_B \rangle$ on $n_F$ in
NSD $pp$ collisions at $\sqrt{s} = 7$~TeV in QGSM. Solid circles and 
solid line denote the distributions for all NSD events and for the 
events without hard processes, respectively. Other symbols indicate 
$\langle n_B (n_F) \rangle$ distributions for sub-events with only soft
Pomerons, $N_{s\_Pom} = 0,1, \ldots, 7$.  
}
\label{fig5}
\end{figure}

Figure~\ref{fig5} displays dependence $\langle n_B \rangle$ on $n_F$ 
for sub-events in NSD $pp$ collisions at $\sqrt{s} = 7$~TeV without hard 
Pomerons and with fixed number of soft Pomerons, increasing from zero 
to seven. All but double-diffractive distributions are remarkably flat 
indicating no forward-backward multiplicity correlations within each
class of selected events. However, the integrated samples of NSD 
collisions with and without the hard processes demonstrate clearly
the positive dependence of $\langle n_B \rangle$ on $n_F$. Positive
slope of full $\langle n_B \rangle(n_F)$ distribution confirms the 
statement that mixing of events with different mean values will lead 
to the correlations in the whole sample \cite{dpm,WS_11}. Compared to
lower energies, hard processes are more abundant at LHC, and their 
contribution makes the slope of $\langle n_B (n_F) \rangle$ steeper.
The non-zero slope of $\langle n_B (n_F) \rangle$ for double-diffractive
processes is also explained by mixing of sub-processes with different
mean multiplicities, see Fig.~\ref{fig1}(e)-(g).

\begin{figure}[htb]
\vspace*{-1.0cm}
 \resizebox{\linewidth}{!}{
\includegraphics{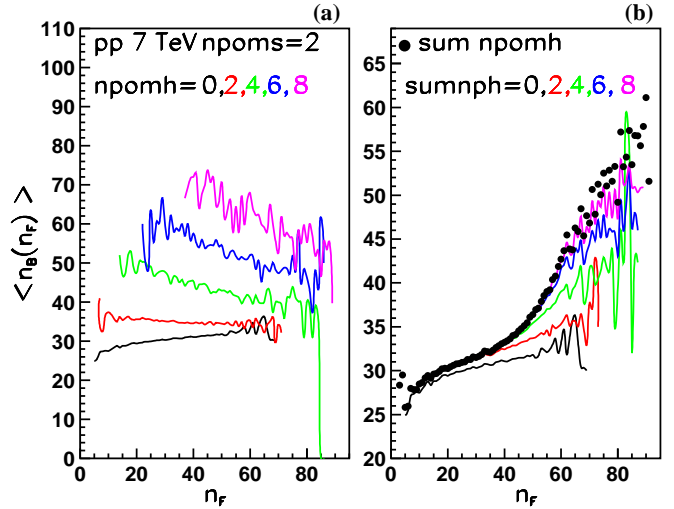}
}
\caption{
The same as Fig.~\ref{fig5} but for sub-events with two soft Pomerons
and with $N_{h\_Pom} = 0, 2, \ldots, 8$ hard Pomerons. Solid symbols show
the $\langle n_B (n_F) \rangle$ dependence for all sub-events with
two soft Pomerons and all hard Pomerons.  
}
\label{fig6}
\end{figure}

\begin{figure}[htb]
\vspace*{-1.0cm}
\includegraphics[scale=0.52]{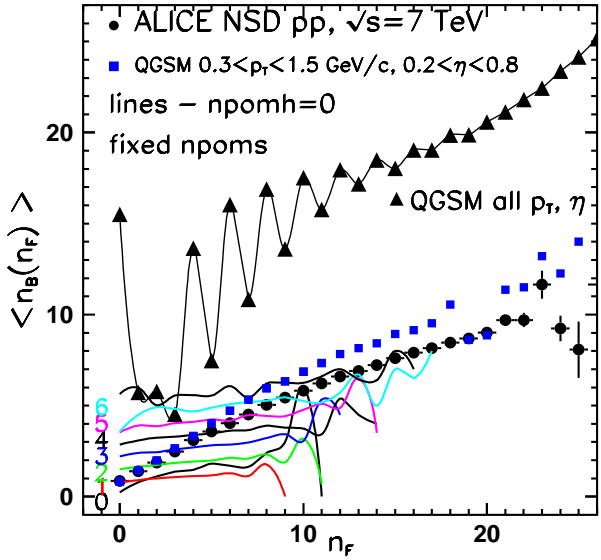}
\vspace*{-1.0cm}
\caption{
The same as Fig.~\ref{fig5} but for hadrons within the kinematic region
$0.3 \leq p_T \leq 1.5$~GeV/$c$ and $0.2 \leq \eta \leq 0.8$. Solid
squares and circles indicate QGSM calculations and ALICE data from
\cite{fb_alice_15}, respectively. Solid triangles denote the 
calculations of NSD events in full phase space.
}
\label{fig7}
\end{figure}

For the events with both soft and hard Pomerons the picture is more
peculiar. Dependencies $\langle n_B (n_F) \rangle$ for sub-events with 
two soft Pomerons and even number of hard Pomerons are shown in
Fig.~\ref{fig6}(a), whereas Fig.~\ref{fig6}(b) depicts the combination 
of spectra with 2,4, etc. hard Pomerons, presented in Fig.~\ref{fig6}(a). 
With the increase of number of hard Pomerons in a sub-event, the 
individual distributions start to develop slightly negative 
slopes. But mixing up these sub-events leads to appearance of positive 
correlation clearly seen in Fig.~\ref{fig6}(b). Therefore, the slope 
$b_{corr}$ increases in the model with rising collision energy because 
more sub-processes with different mean multiplicities become available.

Recall, that the FB correlations are usually explained by the interplay
of the short range correlations arising from the decay of individual
sources and the long range correlations originating from the
fluctuations in the number of sources. Our findings strongly support
this conclusion. By selecting certain sub-events, we fix the number of
strings, i.e., sources of particles. Fluctuations in the number of
sources means mixing of sub-events with different amount of strings
which also have different mean multiplicities.

As was shown in \cite{Am_94}, inclusion of string-string interactions in 
the model via the string fusion mechanism reduces the number of strings 
which leads to decrease of the correlation strength $b_{corr}$. However, 
the reduction of the $b_{corr}$ in $pp$ interactions even at 
ultrarelativistic energies appears to be less than 5\% because of not 
very high densities of the strings, in stark contrast to heavy ion 
collisions \cite{Am_94}.

\subsection{Comparison with the experimental data}
\label{subsec3.2}
    
Average values of $\langle n_B(n_F) \rangle$ in NSD $pp$ collisions at
$\sqrt{s} = 7$~TeV reported in \cite{fb_alice_15} were obtained under 
the kinematic cuts $0.3 \leq p_T \leq 1.5$~GeV/$c$ and $0 \leq \eta \leq 
0.8$. Model calculations shown in Fig.~\ref{fig7} agree reasonably well 
with the data. The average multiplicities $\langle n_B \rangle$ and 
accessible multiplicity intervals 
in $n_F$ for the processes with $n = 1,\ldots,7$ soft Pomerons indicate 
that all sub-processes contribute to events with charged particle 
multiplicity $n_F$ below 10, while for higher $n_F$ the number of 
contributed topologies is reduced. As a consequence, the slope of 
$\langle n_B(n_F) \rangle$ becomes more acclivous.
Note, however, that the imposed kinematic cuts completely remove
the contributions from double-diffractive processes in addition to the
single-diffractive ones. To show this we plot in Fig.~\ref{fig7} QGSM
calculations for NSD $pp$ events in full range of rapidity and 
transverse momentum.
One can see characteristic oscillations of $\langle n_B (n_F) \rangle$
distribution at $n_F \leq 16$, arising in diffractive events because of 
the charge conservation. Namely, the mid-rapidity gap is large and the
most typical configurations are odd numbers of charged particles in each
hemisphere. For even number of charged hadrons in forward hemisphere
there should be emission of one particle from backward hemisphere.
The process is likely to occur in high-multiplicity diffraction after
the decays of resonances. The odd multiplicities in low $n_F$ region
do not need this constraint. Therefore, for low number of $n_F$ the
oscillations in FB correlations are quite significant, whereas the 
larger the number $n_F$, the less pronounced the peaks are.   

\begin{figure}[htb]
 \resizebox{\linewidth}{!}{
\includegraphics{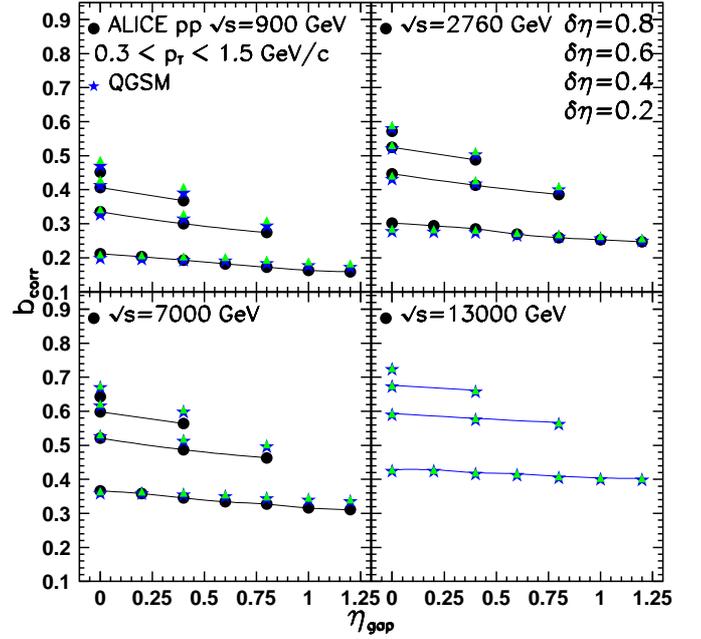}
}
\caption{
FB correlation parameter $b_{corr}$ as a function of $\eta_{gap}$ in
four rapidity bins (from bottom to top): $\delta \eta = 0.2, 0.4, 0.6,
0.8$ in $pp$ collisions at $\sqrt{s} = 900$~GeV, 2.76~TeV, 7~TeV and 
13~TeV. Stars and circles denote the model calculations and the ALICE 
data from \cite{fb_alice_15}, respectively. Lines are drawn to guide 
the eye.
}
\label{fig8}
\end{figure}

\begin{figure}[htb]
 \resizebox{\linewidth}{!}{
\includegraphics{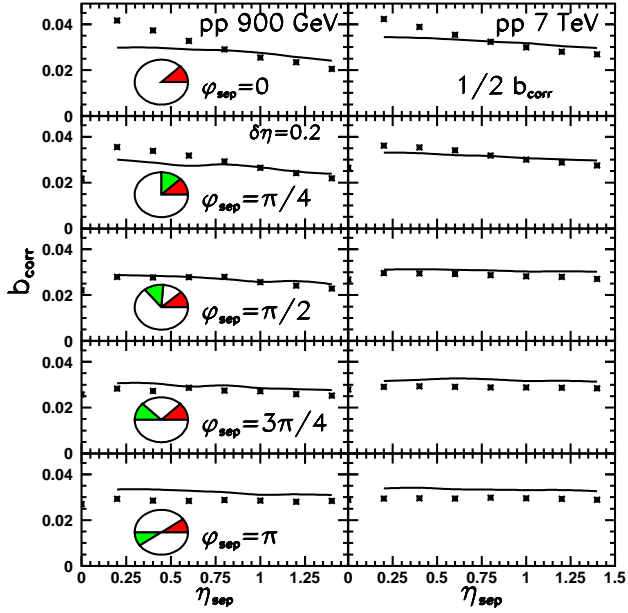}
}
\caption{
FB correlation parameter $b_{corr}$ for azimuthally separated sectors
of $\varphi = \pi/4$ as a function of midrapidity gap at fixed $\delta
\eta = 0.2$ in $pp$ collisions at $\sqrt{s} = 900$~GeV (left column) 
and 7~TeV (right column). Results for 7~TeV are reduced by factor 2.
Lines and asterisks denote the model calculations and the data from
\cite{fb_alice_15}, respectively.
}
\label{fig9}
\end{figure}

\begin{figure}[htb]
 \resizebox{\linewidth}{!}{
\includegraphics{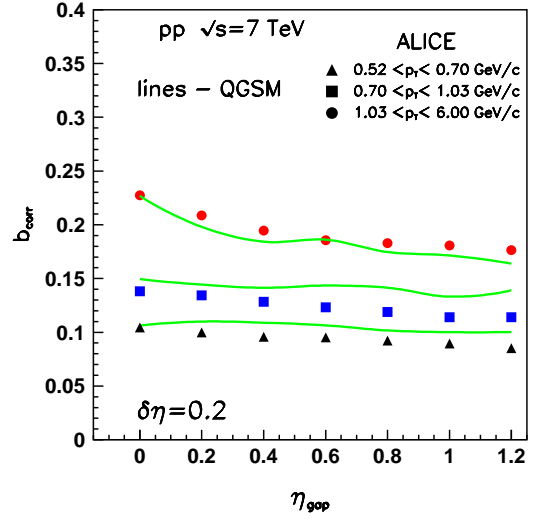}
}
\caption{
FB correlation parameter $b_{corr}$ for windows of width $\delta \eta = 
0.2$ as a function of $\eta_{gap}$ in three $p_T$ intervals:
$0.52\ {\rm GeV}/c \leq p_T < 0.70\ {\rm GeV}/c$ (bottom curve),
$0.70\ {\rm GeV}/c \leq p_T < 1.03\ {\rm GeV}/c$ (medium curve), and
$1.03\ {\rm GeV}/c \leq p_T < 6.0 \ {\rm GeV}/c$ (upper curve) in
QGSM calculations of $pp$ collisions at $\sqrt{s} = 7$\, TeV. Symbols
denote the data from \cite{fb_alice_15}.
}
\label{fig10}
\end{figure}

Further analysis is done in terms of midrapidity gaps $\eta_{gap}$,
hadrons of which are excluded from the investigation, and the widths of 
selected pseudorapidity bins $\delta \eta$ in both hemispheres. 
Following \cite{fb_alice_15}, we vary the $\eta_{gap}$ from 0 to 1.2 
and perform the study for $\delta \eta = 0.2, 0.4, 0.6$ and 0.8. These 
distributions are displayed in Fig.~\ref{fig8} for NSD $pp$ collisions
at $\sqrt{s} = 900$~GeV, 2.76~TeV, 7~TeV and 13~TeV. Comparing the model
results with the data obtained at first three energies, one can see that 
QGSM both quantitatively and qualitatively describes two general 
features. Firstly, parameter $b_{corr}$ increases with 
broadening of $\delta \eta$ at fixed rapidity gap $\eta_{gap}$.
Secondly,  for fixed $\delta \eta$ the strength of FB multiplicity 
correlations drops with increasing midrapidity gap $\eta_{gap}$ in
accord with the observations at lower energies \cite{ua5_2,fb_qgsm_89}. 
This means that the string processes contribute to FB correlations in
central rapidity region, whereas for well-separated areas in forward
and backward hemispheres the long-range FB correlations are absent.
Predictions made for correlations at $\sqrt{s} = 13$~TeV show
further increase of $b_{corr}$. The decrease of the slope 
parameter at fixed $\delta \eta$ with rising $\eta_{gap}$ is not so
steep because of the broadening of midrapidity region.

The FB multiplicity correlations are studied in different azimuthal 
sectors for $pp$ collisions at 900~GeV and 7~TeV. The model calculations 
are plotted onto the ALICE data in Fig.~\ref{fig9}. The width of the 
rapidity bin is $\delta \eta = 0.2$ and the azimuthal angle of each 
sector is $\varphi = \pi/4$. Similar to analysis shown in 
Fig.~\ref{fig8}, the midrapidity gap varies from $\eta_{gap} = 0.2$ to 
1.4. Note, that the results at 7~TeV are reduced by factor 2. Except of
the absolute strength of the correlations, there is a very weak 
difference between the distributions at 900~GeV and 7~TeV. 
The correlation strength between the different sectors in 
Fig.~\ref{fig9} is order of magnitude weaker compared to azimuthally 
integrated results shown in Fig.~\ref{fig8}. It is worth noting the
different $\varphi$-dependence of the FB correlations at midrapidity and 
in more peripheral regions. At $\eta_{gap} = 0.2$ the strength of the 
correlations drops slightly with $\varphi_{sep}$ rising from zero to 
$\pi$, whereas at $\eta_{gap} = 1.2$ a weak increase of the $b_{corr}$ 
is observed.
QGSM reproduces the correlations in $\eta - \varphi$ windows at both 
energies quite well. The 20\% underestimation of the near-side rise of 
the correlation strength at small $\Delta \eta$ can be attributed to 
relative lack of resonances in the model.

Finally, Fig.~\ref{fig10} displays the correlation parameter $b_{corr}$ 
for windows of width $\delta\eta = 0.2$ as a function of $\eta_{gap}$ in
different transverse momentum intervals in $pp$ collisions at $\sqrt{s} 
= 7$\,TeV. Data obtained by ALICE Collaboration \cite{fb_alice_15} are
plotted onto the QGSM calculations as well. We see that $b_{corr}$ 
increases for hadrons with larger $p_T$ for all intervals $\eta_{gap}$.
These data help us to discriminate between the possible scenarios of jet
distribution in pseudorapidity. If the jets are uniformly distributed,
the $p_T$ dependence of the $b_{corr}$ in QGSM is similar to that 
obtained in PHOJET model (see Fig.~13b of \cite{fb_alice_15}). Correct
description of the data, shown in Fig.~\ref{fig10}, is attained if one
assumes the Gaussian distribution of jets in $\eta$-space.   

\section{Conclusions}
\label{sec4}

Multiplicity correlations in forward and backward hemispheres are 
studied in $pp$ collisions at energies $200~{\rm GeV}~ \leq \sqrt{s} 
\leq 13~{\rm TeV}$ within the quark-gluon string model. No tuning of
the free parameters of the model was performed. The linear dependence of 
$\langle n_B \rangle $ on $n_F$ is reproduced within the whole energy
range. The main contribution to the FB correlations comes from the 
multistring processes due to multi-Pomeron exchanges. These 
processes have different multiplicity distributions, and the FB 
multiplicity correlations arise because of superposition of such 
sub-processes with different mean multiplicities. For the events with 
fixed amount of soft and hard Pomerons the forward-backward correlations 
are absent. The increase of the variety of sub-processes explains the 
rise of the correlation strength $b_{corr}$ with increasing collision 
energy. Deviation of the $\langle n_B(n_F) \rangle $ distributions from 
linear behaviour in the range of high $n_F$ multiplicities is due to
reduction of number of multi-particle processes contributing to these 
events.

Comparison with experimental data shows that QGSM correctly reproduces
(i) increase of $b_{corr}$ with rising $\sqrt{s}$; (ii) increase of
$b_{corr}$ with broadening width of the bin $\delta \eta$ at 
midrapidity range; (iii) decrease of $b_{corr}$ with increasing 
(mid)rapidity gap $\Delta \eta$ between the tested FB particle samples;
and (iv) FB multiplicity correlations in $\eta - \varphi$ windows.
Experimental data on changing of $b_{corr}(\eta_{gap})$ with rising 
transverse momentum favour the Gaussian distribution of jets in 
pseudorapidity.
Predictions are made for $\sqrt{s} = 13$~TeV. Also, we predict 
oscillations in strength of FB correlations in both single-diffractive 
and double-diffractive collisions in the full phase space. In these 
interactions the distributions $\langle n_B(n_F) \rangle $ should peak 
at even values of $n_F$ and drop at odd $n_F$ values for $n_F \leq 16$. 
The origin of the oscillations is linked to the conservation of electric 
charge. 

{\bf Acknowledgments.}
The authors are grateful to I. Altsybeev for consultations concerning
the analysis of ALICE data. This work was supported by the Norwegian 
Research Council (NFR) under grant No. 255253/F50 $-$ CERN Heavy Ion 
Theory. J.B. thanks the German Research Foundation (DFG) for the 
financial support through the Project BL~1286/2-1. L.B. acknowledges 
financial support of the Alexander von Humboldt Foundation.



\begin{thebibliography}{99}


\bibitem{WDK_96} E.A.~De~Wolf, I.M.~Dremin, W.~Kittel,
Phys. Rep. 270 (1996) 1.

\bibitem{Koch_10} V.~Koch,
in {\it Relativistic Heavy Ion Physics\/}, Landolt-B{\"o}rnstein
Database Vol. 23, ed. R.~Stock (Springer, Berlin, 2010), p.626$-$652.

\bibitem{ua5_1} K.~Alpgaard, et al., UA5 Collab.,  
Phys. Lett. B 123 (1983) 361.

\bibitem{ua5_2} G.G.~Alner, et al., UA5 Collab.,  
Phys. Rep. 154 (1987) 247.

\bibitem{ua5_3} R.E.~Ansorge, et al., UA5 Collab.,  
Z. Phys. C 37 (1988) 191.

\bibitem{uhlig_78} S.~Uhlig, I.~Derado, R.~Meinke, H.~Preissner,
Nucl. Phys. B 132 (1978) 15.

\bibitem{fb_qgsm_89} L.V.~Bravina, et al., 
Sov. J. Nucl. Phys. 50 (1989) 245.

\bibitem{na22_89} V.V.~Aivazyan, et al., NA22 Collab.,
Z. Phys. C 42 (1989) 533.

\bibitem{e735} T.~Alexopoulos, et al., E735 Collab.,
Phys. Lett. B 353 (1995) 155. 

\bibitem{fb_alice_15} J.~Adam, et al., ALICE Collab.,
J. High Energy Phys. JHEP05 (2015) 097.

\bibitem{tasso_89} W.~Braunschweig, et al., TASSO Collab.,
Z. Phys. C 45 (1989) 193.

\bibitem{opal_94} R.~Akers, et al., OPAL Collab., 
Phys. Lett. B 320 (1994) 417.

\bibitem{CK_78} A.~Capella, A.~Krzywicki, Phys. Rev. D 18 (1978) 4120.

\bibitem{CV_82} A.~Capella, J.Tran Thanh Van, Z. Phys. C 18 (1983) 85.

\bibitem{DdD_81} J.~Dias~de~Deus, Phys. Lett. B 100 (1981) 177.

\bibitem{CY_84} T.T.~Chou, C.N.~Yang, Phys. Lett. B 135 (1984) 175.

\bibitem{Car_88} B.~Carazza, Nuovo Cim. 99 (1988) 731.

\bibitem{Bar_87} S.~Barshay, Phys. Lett. B 199 (1987) 121.

\bibitem{CKC_97} L.K.~Chen, D.~Kiang, C.K.~Chew,
Phys. Lett. B 408 (1997) 422.

\bibitem{Am_94} N.S.~Amelin, et al.,
Phys. Rev. Lett. 73 (1994) 2813.

\bibitem{P_qm99} N.~Armesto, M.A.~Braun, E.G.~Ferreiro, C.~Pajares,
Nucl. Phys. A 661 (1999) 325c.

\bibitem{BPV_00} M.A.~Braun, C.~Pajares, V.V.~Vechernin,
Phys. Lett. B 493 (2000) 54.

\bibitem{Vec_15} V.V.~Vechernin, Nucl. Phys. A 939 (2015) 21.

\bibitem{ALICE_ridge} K.~Aamodt {\it et al.} (ALICE Collaboration),
Phys. Rev. Lett. 107 (2011) 032301.

\bibitem{ATLAS_ridge} G.~Aad {\it et al.} (ATLAS Collaboration), 
Phys. Rev. C 86 (2012) 014907.

\bibitem{proton_ridge} V.~Khachatryan {\it et al.} (CMS Collaboration),
JHEP 1009 (2010) 091.

\bibitem{qgsm_1} A.B.~Kaidalov, Phys. Lett. B 116 (1982) 459.

\bibitem{qgsm_2}
A.B.~Kaidalov, K.A.~Ter-Martirosyan, Phys. Lett. B 117 (1982) 247.

\bibitem{qgsm_3}
A.B.~Kaidalov, Phys. Usp. 46 (2003) 1121.

\bibitem{qgsm_mc1} 
N.S.~Amelin, L.V.~Bravina, Sov. J. Nucl. Phys. 51 (1990) 133.

\bibitem{qgsm_mc2}
N.S.~Amelin, L.V.~Bravina, L.I.~Sarycheva, L.N.~Smirnova,
Sov. J. Nucl. Phys. 51 (1990) 535. 

\bibitem{dpm} A.~Capella, U.~Sukhatme, C.I.~Tan, J.~Tran Thanh Van,
Phys. Rep. 236 (1994) 225. 

\bibitem{paciae_10}
Y.-L.~Yan, et al., 
Nucl. Phys. A 834 (2010) 320c.

\bibitem{WS_11} K.~Wraight, P.~Skands, Eur. Phys. J. C 71 (2011) 1628.

\bibitem{KV_14} V.~Kovalenko, V.~Vechernin, 
DESY Conf. Proc. 2014-04/82 (2014) 691 (arXiv:1410.3884 [hep-ph]). 


\bibitem{tH_74} G.~t'Hooft, Nucl. Phys. B 75 (1974) 461.

\bibitem{Ven_74} G.~Veneziano, Phys. Lett. B 52 (1974) 220.

\bibitem{RFT_1} V.~Gribov, Sov. Phys. JETP 26 (1968) 414.

\bibitem{RFT_2}
L.V.~Gribov, E.M.~Levin, M.G.~Ryskin, Phys. Rep. 100 (1983) 1.

\bibitem{ASC_92} N.S.~Amelin, E.F.~Staubo, L.P.~Csernai,
Phys. Rev. D 46 (1992) 4873.

\bibitem{FO_sps} 
L.V.~Bravina, et al., 
Phys. Rev. C 60 (1999) 044905.

\bibitem{v2_plb_01} 
E.E.~Zabrodin, C.~Fuchs, L.V.~Bravina, Amand~Faessler,
Phys. Lett. B 508 (2001) 184.
 
\bibitem{prd_16} J.~Bleibel, L.V.~Bravina, E.E.~Zabrodin,
Phys. Rev. D 93 (2016) 114012.

\bibitem{KP_11} A.B.~Kaidalov, M.G.~Poghosyan,
Eur. Phys. J. C 67 (2010) 397.

\bibitem{epos} K.~Werner, F.-M.~Liu, T.~Pierog,
Phys. Rev. C 74 (2006) 044902.

\bibitem{phojet} R.~Engel, J.~Ranft, S.~Roesler,
Phys. Rev. D 52 (1995) 1459.

\bibitem{qgsjet2} S.~Ostapchenko,
Nucl. Phys., Proc. Suppl. 151 (2006) 143;
Phys. Rev. D 83 (2011) 014018.

\bibitem{AGK} V.~Abramovskii, V.~Gribov, O.~Kancheli,
Sov. J. Nucl. Phys. 18 (1974) 308.

\bibitem{CTVK_87} A.~Capella, J.~Tran~Thanh~Van, J.~Kweicinski,
Phys. Rev. Lett. 58 (1987) 2015.

\bibitem{enh_diag_1} O.V.~Kancheli, JETP Lett. 11 (1970) 267.

\bibitem{enh_diag_2} A.H.~Mueller, Phys. Rev. D 2 (1970) 2963.

\bibitem{AGK-viol1} Y.~V.~Kovchegov, K.~Tuchin,
Phys. Rev. D 65 (2002) 074026.

\bibitem{AGK-viol2} E.~Levin, A.~Prygarin,
Phys. Rev. C 78 (2008) 065202.

\bibitem{FF_frag} R.D.~Field, R.P.~Feynman,
Nucl. Phys. B 136 (1978) 1.

\end{thebibliography}
\end{document}